# Orientation of the intra-unit-cell magnetic moment in the high-$T_c$ superconductor $HgBa_2CuO_{4+\delta}$

Yang Tang[1], Lucile Mangin-Thro[2], Andrew Wildes[2], Mun K. Chan[1], Chelsey J. Dorow[1], Jaehong Jeong[3], Yvan Sidis[3], Martin Greven[1,*], Philippe Bourges[3,*]

[1] *School of Physics and Astronomy, University of Minnesota, Minneapolis, MN 55455, USA*

[2] *Institut Laue-Langevin, 71 avenue des martyrs, Grenoble 38000, France*

[3] *Laboratoire Léon Brillouin, CEA-CNRS, Université Paris-Saclay CEA-Saclay, Gif sur Yvette 91191, France*

\* greven@umn.edu, philippe.bourges@cea.fr

**Abstract**

Polarized-neutron diffraction experiments (PND) have revealed that the pseudogap state of the cuprates exhibits unusual intra-unit-cell (IUC) magnetism. At a qualitative level, the data indicate a moment direction that is neither perpendicular nor parallel to the $CuO_2$ layers, yet an accurate measurement of a structurally simple compound has been lacking. Here we report PND results with unprecedented accuracy for the IUC magnetic order in the simple-tetragonal single-$CuO_2$-layer compound $HgBa_2CuO_{4+\delta}$. At the transition temperature, we find evidence for magnetic critical scattering. Deep in the ordered state, we determine the moment direction to be $70° \pm 10°$ away from the normal to the $CuO_2$ layers, which rules out both purely planar loop currents and high-symmetry Dirac multipoles, the two most prominent theoretical proposals for the microscopic origin of the IUC magnetism. However, the data are consistent with Dirac multipoles of lower symmetry or, alternatively, with a particular configuration of loop currents that flow on the faces of the $CuO_6$ octahedra.

**Introduction**

The lamellar high-temperature superconducting cuprates exhibit unusual properties as a result of their strong quasi-two-dimensional electronic correlations. One of the most interesting characteristics of these complex oxides is the pseudogap phenomenon, whose origin has been under intense debate [1]. Numerous experiments indicate that the pseudogap state is a distinct phase of matter, including circularly-polarized photoemission [2], polarized-neutron diffraction [3-12], polar Kerr effect [13], resonant ultrasound [14], optical birefringence [15], second-harmonic-generation optical response [16], torque magnetometry [17], and muon spin relaxation (µSR) [18]. The PND experiments span four different cuprate families and point to unusual IUC magnetic order (reduced wave vector $\mathbf{q} = 0$) that preserves the lattice translational symmetry. The IUC magnetic signal is observed below a doping-dependent characteristic temperature (denoted as either $T_{q=0}$ or $T_{mag}$) that matches the characteristic pseudogap temperature ($T^*$) determined from planar resistivity measurements (Fig. 1a). This demonstrates that the IUC magnetic order is one of the hallmarks of the pseudogap phase. In the underdoped part of the phase diagram, the IUC magnetic order precedes the superconductivity and other electronic instabilities, such as charge-density-wave order (Fig. 1a).

Since the IUC magnetic order does not produce a net magnetization, it can be naively thought of as a simple superposition of an even number of moments that cancel out within each primitive cell. A state that gives rise to such magnetism was actually theoretically predicted [19,20] prior to the experimental findings. In this 'loop-current' (LC) model, spontaneous LCs develop within each square Cu-O plaquette. Orbital magnetism may arise from either two or four counter-circulating LCs per plaquette. The PND data are qualitatively consistent with the two-LC scenario [4-12], which is also supported by variational Monte Carlo calculations [21]. Whereas in the original model the orbital moments point perpendicular to the $CuO_2$ planes, the PND data indicate a significant in-plane component, albeit with rather large experimental uncertainty [3, 5, 6, 9, 10]. In a revised version of the original planar LC model, it was argued that such a magnetic signal might originate from a quantum superposition of (classical) LC patterns [22]. Alternatively, the LCs might flow on the faces of the $CuO_6$ octahedra that surround Cu sites in a single-layer material such as $HgBa_2CuO_{4+\delta}$ (Hg1201), or on the faces of the $CuO_5$ pyramids in double-$CuO_2$-layer compounds such as $YBa_2Cu_3O_{6+x}$ (YBCO) [4, 23-25]. Two distinctly different microscopic pictures involve planar oxygen moments [3] and Dirac (or magneto-electric) multipoles [26-28]. In order to help distinguish among these scenarios it therefore is of considerable importance to determine the orientation of the IUC moments with higher precision. This would be best achieved in a structurally simple cuprate compound.

Whereas the microscopic nature of the IUC magnetism remains an open question, its existence has been firmly established through PND experiments performed on four different cuprate families. Perhaps the most important theoretical question is the relation between the IUC order and the pseudogap. The original LC model [21,22] faces a problem, since it can explain the IUC/**q**=0 order reported by various measurements, but not the opening of the pseudogap. However, it has been argued that this problem is circumvented if the order is not truly long-range [29]. On the other hand, it has been argued that topological order can open a pseudogap and give rise to an emergent LC phase with a symmetry consistent with the neutron experiments [30]. Furthermore, various models imply charge- or pair-density-wave instabilities, e.g., with a composite *d*-wave superconducting and charge-density wave with emergent SU(2) symmetry [31, 32]. In these models, the pseudogap can be viewed as a phase of fluctuating superconducting correlations, and $T^*$ may be a crossover temperature. However, a preemptive phase that breaks both parity and time-reversal symmetry, such as the original LC phase, is expected at $T^*$ [31]. A very recent proposal, which is based on the experimental facts that the pseudogap is spatially inhomogenous [33] and that no large thermodynamic anomaly is observed at $T^*$ [34], argues that the pseudogap formation is a percolative phenomenon associated with gradual inhomogeneous charge localization [35]. In this scenario, the IUC order is an emergent phenomenon that does not significantly affect the electronic density of states at the Fermi level. In these latter models, IUC order is an important ingredient and occurs systematically at higher temperature prior to subsequent instabilities. In the present paper, we wish to better characterize the IUC order in a model experimental system, with particular focus on the question how to describe this state in terms of either LCs or magnetic multipoles.

We report PND measurements for two Hg1201 samples (Fig. 1a), one moderately under-doped (superconducting transition temperature $T_c$ = 71 K, hole doping level $p \approx 0.095$; denoted UD71) and one nearly optimally-doped ($T_c$ = 95 K, $p \approx 0.127$; OP95). The use of polarized neutrons is required in order to discern relatively weak magnetic Bragg signal from strong underlying nuclear Bragg diffraction. Prior measurements on samples grown by the same method [36], such as X-ray scattering [37], charge transport [38-40], optical spectroscopy [41], and inelastic neutron scattering [42,43] indicate that Hg1201 can be considered a model cuprate compound. Neutron diffraction results for Hg1201 [4,7] are highly consistent with the original discovery of the IUC magnetic order in YBCO [3,6]. Hg1201 has a particularly simple structure (high tetragonal P4/mmm crystal symmetry, one $CuO_2$ layer per primitive cell, no Cu-O chains), exhibits relatively small disorder effects [38-40], and features an optimal $T_c$ of about 97 K, the highest among all single-layer cuprates. Hg1201 thus is a very promising compound for the study of the pseudogap magnetism. Unlike the previous PND studies of Hg1201, which focused on (1 0 $L$) reflections with nonzero integer $L$ [4,7], we choose the high-symmetry reflection (1 0 0) in the present work, as this enables improved polarization analysis. In particular, any wave vector $\mathbf{Q} = (H\ K\ L)$ with nonzero out-of-plane component $L$ results in the measurement of a superposition of in-plane and out-of-plane magnetic moments, rendering them difficult to distinguish in the polarization analysis. Moreover, ($H$ 0 0)-type reflections have a unit-cell structure factor for magnetic neutron diffraction that is identically zero for axial dipoles and uniquely sensitive to Dirac multipoles [28].

**Results**

In contrast to the previous reports of IUC magnetic order in Hg1201 [4,7], we perform measurements on a different (multi-detector) diffractometer (D7, at the Institute Laue-Langevin, Grenoble, see *Methods*) [11]. Genuine magnetic scattering can be obtained through longitudinal polarization analysis using the classic XYZ-polarization analysis technique [3,6,8,11]. The Hg1201 samples were mounted in the ($H$ 0 $L$) horizontal scattering plane. We observe magnetic signal only for (1 0 $L$)-type Bragg reflections, consistent with the prior work [4,7] (see supplementary Figure S4). Figure 2 shows the temperature dependence of the inverse of the flipping ratio, 1/*FR*, for both UD71 and OP95. 1/*FR* is defined as the ratio of the measured spin-flip (SF) intensity to the measured non-spin-flip (NSF) intensity, and $i = \{X,Y,Z\}$ denotes the three neutron polarizations: $1/FR_i = I_i^{SF} / I_i^{NSF}$. For OP95, $1/FR_i(T)$ decreases in a gradual, monotonic fashion with decreasing temperature and can be described (below about 400 K) by a polynomial fit. This behaviour is consistent with the lack of any magnetic Bragg signal, which would be expected to lead to an increase with decreasing temperature. In contrast, $1/FR_i(T)$ for UD71 exhibits an upturn below $T_{q=0}$ = 360 - 380 K. A magnetic signal is thus observed for UD71 and absent (or very small) for OP95, in agreement with prior observations [5,8]. Furthermore, 1/$FR_i$(T) for UD71 noticeably depends on the polarization, with a maximum amplitude in Y-polarization.

In order to describe this polarization dependence, we first decompose the IUC magnetic moment ***m*** into the three polarization directions. The magnetic moment is a superposition of moments along the reciprocal lattice basis, $m^2 = m_a^2 + m_b^2 + m_c^2$. Since Hg1201 has tetragonal symmetry, and hence ***a**** and ***b**** are equivalent, the in-plane magnetic components are equal: $m_a^2 = m_b^2$. We therefore can simply express the moment in terms of the in-plane component ***m**$_{ab}$*, with $m_{ab}^2 = 2m_a^2$, and out-of-plane component ***m**$_c$*, which are related to the total magnetic moment as $m^2 = m_{ab}^2 + m_c^2$. We define ϕ to be the angle between the ***c****-axis and the total magnetic moment ***m***: *tan(ϕ)=m$_{ab}$/m$_c$* (see Fig. 1b). The magnetic components are related to the magnetic contributions $M_i$ along the three polarization directions. In the SF channel, for **Q** = (1 0 0) [12]:

$$M_Z \propto m_c^2 \tag{1}$$

$$M_Y \propto \frac{1}{2}m_{ab}^2 + sin^2\alpha\, m_c^2 \tag{2}$$

$$M_X \propto \frac{1}{2}m_{ab}^2 + cos^2\alpha\, m_c^2 \tag{3}$$

where α is defined as the angle between the momentum transfer ***Q*** and the polarization direction X (Fig. S1b). In the limit where α = 0, the magnetic intensity follows the sum-rule $M_X = M_Y + M_Z$ discussed in our previous reports using longitudinal polarization analysis on triple-axis spectrometers [3-10]. On the diffractometer D7, $\alpha = 108.2° \pm 5°$ for ***Q*** = (1 0 0) (see Supplementary Information). The relations (1) - (3), which are specifically satisfied for magnetic scattering, show that the magnetic signal should be maximum in Y polarization, as is indeed observed for UD71 (Fig. 2).

Prior PND work revealed an order-parameter-like temperature dependence [5,6] for the **q** = 0 magnetic moment, and we therefore write:

$$m_{ab,c}(T) = m_{ab,c}\left(1 - \frac{T}{T_{q=0}}\right)^\beta \tag{4}$$

with $T_{q=0}$ the onset of the **q** = 0 order and β the effective exponent that describes the observed power-law-like temperature dependence. The detailed, quantitative data analysis to extract the **q** = 0 magnetic signal is described the Supplementary Information. The data are analysed in two different ways, which both yield essentially the same result. Method 1 assumes that both UD71 and OP95 exhibit **q** = 0 magnetism on top of a SF background with the same linear temperature dependence. The lines in Fig. 3 are obtained from these fits. For UD71, we find $T_{q=0} = 370 \pm 30$ K, slightly higher than the characteristic temperature *T\** obtained from planar resistivity measurements [33,34] (Fig. 1), and $\beta \approx 0.25 \pm 0.05$, consistent with the prior data for Hg1201 [7] and YBCO [6]. The extracted values (obtained in arbitrary units; see also Table 1) of $m_{ab}^2$ and $m_c^2$ are $0.58 \pm 0.14$ a.u. and $0.20 \pm 0.06$ a.u., respectively, which corresponds to $\phi = 71° \pm 10°$. For OP95, on the other hand, the in-plane and out-of-plane moments are zero within error, with an upper limit of about 0.08 a.u. for both. This absence of a discernible magnetic Bragg signal is consistent with the previously reported result for a nearly optimally doped Hg1201 sample ($T_c$ = 89 K, $p \approx 0.116$) [5].

One can estimate the signal strength in absolute units from a comparison with the intensity of the (1 0 0) nuclear Bragg peak, which is ~ 300 a.u. and calculated to be 4.6 barn based on the composition and crystal structure of Hg1201. Then, assuming that the magnetic signal is long-range, the total magnetic intensity, $m^2 = m_{ab}^2 + m_c^2$ is found to be $9.4 \pm 2$ mbarn for UD71, consistent with previous estimation [4]. This results in an upper bound of ~1.7 mbarn for OP95 at **Q** = (1 0 0) (consistent with complementary triple-axis data, shown in Fig. 6S).

Figure 3 shows the magnetic intensity obtained by Method 2, where we assume no discernible magnetic signal in OP95 and use this as a background reference for UD71 (see also the Supplementary Information). Importantly, the signal satisfies polarization analysis, which demonstrates its magnetic origin: the solid lines in Fig. 3 show the fit to (1) - (4) below $T_{q=0}$ ~ 360 K, with $\beta \approx 0.20 \pm 0.05$. The values for $m_{ab}^2$ and $m_c^2$ are listed in Table 1 together with those obtained from Method 1. The two methods give consistent results, within error. We obtain $T_{q=0} \approx 360 \pm 30$ K for the mean value, consistent with *T\** from planar resistivity measurements [33,34] (Fig. 1a).

Recent µSR measurements of YBa$_2$Cu$_3$O$_{6+x}$ [18] and Bi$_2$Sr$_2$CaCu$_2$O$_{8+\delta}$ [44] revealed slow magnetic fluctuations and critical slowing down in the pseudogap phase. In particular, the µSR longitudinal relaxation rate was found to go through a maximum at the temperature $T_{q=0}$ (or $T_{mag}$), a characteristic of critical slowing down typically associated with a second-order phase transition. In a PND study of nearly-optimally-doped YBCO$_{6.85}$ [10], performed on the D7 diffractometer used in the present work, evidence for critical-like magnetic scattering was reported at **Q** = (0.88 0 0), i.e., off the Bragg position. In Fig. 4b, we reproduce these data by plotting the sum of all SF cross-sections, $\Sigma_{SF} = I_{SF}^X + I_{SF}^Y + I_{SF}^Z$. Consistent with the µSR relaxation rate [18], $\Sigma_{SF}$ exhibits a peak at the onset temperature of the **q** = 0 magnetic order suggestive of critical slowing down. For both UD71 and OP95, the same quantity $\Sigma_{SF}$, determined at **Q** = (0.88 0 -0.11), displays a peak as well at a temperature close to the pseudogap temperature $T^*$ (Fig. 4a). This feature can also be seen from the first derivative of $\Sigma_{SF}$ (Fig. S5b), which exhibits a sharp S-shape for both samples at the respective $T^*$. For OP95, a complementary measurement of the magnetic scattering, $\Delta_{SF}$, obtained through polarization analysis around **Q** = (0.9 0 0) on the triple-axis spectrometer 4F1 at LLB/Orphée also shows a maximum (near 220 K; see Figure S7).

For UD71, the characteristic temperature of the maximum of $\Sigma_{SF}$ is consistent with the longitudinal polarization analysis at the Bragg position (1 0 0) (Figs. 2-3). For OP95, one can define the temperature $T_{q=0} \sim 200$ K from the anomaly in Fig. 4 although the magnetic intensity was not discernible at the Bragg position (figure 2). In contrast, additional measurement of OP95 at the (1 0 0) and (1 0 1) reflections on 4F1 with a coarser Q-resolution than D7 revealed evidence for a magnetic scattering at the Bragg positions (Fig. S6 and supplementary Material S4). Therefore, in light of the observation for nearly optimally-doped YBCO$_{6.85}$ of short-range, rather than long-range magnetic order [11], we propose that the **q** = 0 magnetism in OP95 is short-range as well. As a consequence, the magnetic response is redistributed throughout the Brillouin zone and not discernible at the Bragg peak (1 0 0) within the experimental conditions of the D7 instrument. These observations motivated us to search for a weak magnetic signal away from the (1 0 0) Bragg position. Figure 4c shows the difference between **Q** = (*H* 0 -0.4) momentum scans in the SF channel across *H* = 1 obtained at 150 K (below $T_{q=0}$) and 225 K (above $T_{q=0}$). Indeed, we are able to discern a net magnetic signal at *H* = 1, consistent with the existence of short-range IUC magnetic order in OP95. A rough estimate of the in-plane correlation length yields $\xi/a \sim 5$, a value that is even shorter than for YBCO$_{6.85}$ [11].

**Discussion**

Figures 2-3 demonstrate that the observation of a magnetic signal in the pseudogap state of UD71 is independent of the data analysis method and consistent with a second-order phase transition at $T_{q=0} = 360 \pm 30$ K, accompanied by magnetic critical fluctuations. The current results for Hg1201, obtained on a diffractometer with unprecedented signal-to-noise ratio, confirm prior work which employed a triple-axis spectrometer [3-10]. For OP95, the magnetic

Bragg signal is at least one order of magnitude weaker (see Table 1). Nevertheless, evidence for critical fluctuations near $T_{q=0} = 210 \pm 30$ K (Fig. 4) and short-range correlations is observed. Overall, these estimates are consistent with previous results for Hg1201 [7,12] (Fig. 1b), and with results for YBCO, where a similar evolution from long-range 3D magnetic correlations at low doping toward short-range 2D magnetic correlations near optimal doping was observed [11].

For UD71, quantitative longitudinal polarisation analysis at the (1 0 0) Bragg reflection yields in-plane and out-of-plane IUC magnetic components with unprecedented accuracy (Table 1). The moment direction is tilted away from the crystallographic $c$ axis by $\phi = 70° \pm 10°$. This value is somewhat larger, yet consistent with those obtained for other cuprates (Table 2).

Our results (Table 1) allow us to rule out models with strictly in-plane ($\phi = 90°$) or out-of-plane ($\phi = 0$) IUC moments. Specifically, we can rule out all models where $\phi$ goes to 0 at $L = 0$, in particular the original planar LC model [19,20], and models where $\phi$ goes to 90° at $L = 0$, in particular the magneto-electric multipole scenarios with quadratic symmetry of refs. [26-28] in which the out-of-plane moment component is zero. However, we cannot rule out variations of these scenarios, either within the LC picture, where an in-plane component might appear due to quantum corrections [22], or quadrupolar order with monoclinic symmetry [27,28]. The quadrupole lobes (or current loops) exhibit different spontaneous magnetic fields on opposite sides of a Cu atom. The neutron spin moment probes these different microscopic magnetic patterns, and the interference between them, and the corresponding cross section can be expected to be largest when the size of the quadrupole lobes (or current loops) is comparable to the neutron wavelength. An interesting scenario that might explain our data is the dual existence of planar LC order and magneto-electric quadrupoles, as both can be treated on the same ground [45].

Our result also is consistent with a variant of the LC model in which charge currents flow on the faces of the oxygen pyramids/octahedra (Fig. 4c). For Hg1201, this corresponds to an angle of about 64°, as calculated from lattice parameters [4]. Several variants of this scenario have been considered [4,23-25,46,47]. However, most of these variants are inconsistent with our data at the high symmetry point, $L = 0$. Indeed, structure factor calculations show that the variants considered in [4,23,24,47] exhibit out-of-plane and in-plane components at different Bragg positions: only the out-of-plane component contributes to the (1 0 0) reflection whereas the in-plane component would result in intensity at (0 0 L) Bragg peaks, which has not been observed in experiment. Only the specific variant with two current loops depicted in Fig. 4c (originally considered in ref. [47]) is consistent both with the neutron and Kerr-effect data [13, 48]; in this scenario, the currents flow on opposite faces of the two pyramids that form the $CuO_6$ octahedra of Hg1201.

YBCO features pairs of $CuO_5$ pyramids associated with adjacent $CuO_2$ planes rather than $CuO_6$ octahedra associated with a single plane. The faces of the pyramids form an angle of about 59° [4] with the $CuO_2$ planes. According to earlier results for (twin free) underdoped YBCO [12], the out-of-plane magnetic scattering exhibits an *a-b* anisotropy, which is furthermore *L*-dependent. This feature can be accounted for by a crisscrossed stacking of planar LCs and eliminates as a possible origin of the out-of-plane magnetic scattering all magnetic patterns that do not break parity, such as magnetism on the oxygen sites. Further, the PND data for nearly optimally-doped YBCO show the absence of a tilt ($\phi = 0$) at high temperature, where IUC magnetic correlations develop, and that $\phi$ acquires a nonzero value of $40 \pm 9°$ at $T_{q=0}$ [10]. This variation of $\phi$ as a function of temperature is consistent with a crossover from classical to quantum planar LC correlations [22], with the coexistence of planar LC order and another form of IUC magnetic order (the latter controlling the in-plane magnetic scattering intensity), and also with a crossover from planar to out-of-plane LC order. In the latter two scenarios, the tilt angle reflects the degree of admixture either of different kinds of IUC orders, or of planar and out-of-plane currents, which might change not only with temperature, but also with doping. In this regard, it is interesting to note that $\phi$ is quite large for the underdoped Hg1201 sample, reaching $70 \pm 10°$, whereas, for a nearly optimally-doped $Bi_2Sr_2CaCu_2O_{8+\delta}$ the angle can be as small as $20 \pm 20°$ below $T_{q=0}$ (Table 2).

In conclusion, we have conducted a quantitative polarized-neutron diffraction study of the model cuprate Hg1201. Consistent with prior work, we observe robust **q** = 0 magnetic order in the pseudogap state of a moderately-doped sample with $T_c \approx 71$ K, and evidence for short-range correlations in a nearly optimally-doped sample with $T_c \approx 95$ K. In the former case, analysis of the data obtained at the (1 0 0) reflection yields the estimate $\phi = 70° \pm 10°$ for the tilt direction of the magnetic moment away from *c*-axis. This estimate constitutes a significant improvement over prior data and places new constraints on the microscopic origin of the observed intra-unit-cell magnetism.

**Methods**

Both HgBa$_2$CuO$_{4+\delta}$ samples studied by neutron diffraction are described in the Supplementary Information. The spin-polarized neutron diffraction experiments were performed on the cold neutron diffractometer D7 at the Institute Laue-Langevin, Grenoble, France. The experimental set-up of D7 was similar to that of a previous study of YBa$_2$Cu$_3$O$_{6+x}$ [11] and is described in the Supplementary Information. We quote the scattering wave-vector **Q** = H**a*** + K**b*** + L**c*** ≡ (*H K L*) in reciprocal lattice units, where **a*** = **b*** = 1.62 Å$^{-1}$ and **c*** = 0.66 Å$^{-1}$ are the approximate room-temperature values.

**Acknowledgements:** The work at the University of Minnesota was funded by the Department of Energy through the University of Minnesota Center for Quantum Materials under DE-SC-0016371. We also acknowledge financial support at LLB from the projects UNESCOS (contract ANR-14-CE05-0007) and NirvAna (contract ANR-14-OHRI-0010) of the ANR. We acknowledge decisive discussions with Sergio di Matteo, Thierry Giamarchi, Stephen Lovesey, Mike Norman and Chandra Varma on the topic of this article.

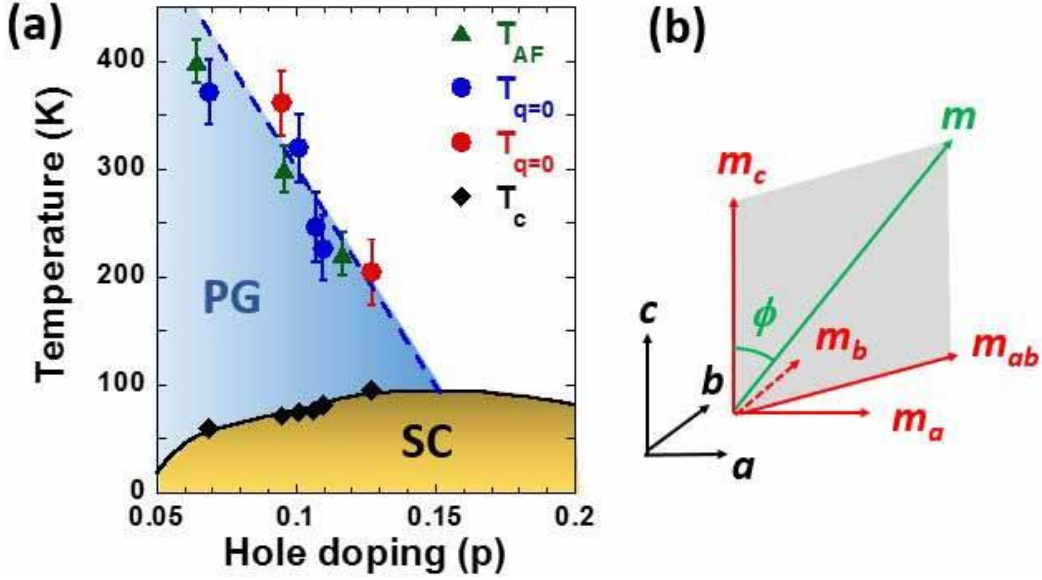

**Figure 1.** (a) Phase diagram of Hg1201. Red symbols pertain to the samples studied in the present work. The superconducting (SC) phase is shown as brown shaded area. The hole doping level ($p$) is determined from the $T_c(p)$ relationship according to [49]. Neutron scattering experiments reveal two characteristic temperatures associated with the pseudogap (PG) phase (light blue area): $T_{q=0}$ and the onset temperature of the antiferromagnetic fluctuations $T_{AF}$ [42, 43]. These temperatures are consistent with the characteristic pseudogap temperature $T^*$ obtained from charge transport measurements [38, 50]. In an intermediate $p$-$T$ range within the PG phase, charge order is reported below $T_{CO}$ ($T_{CO} < T^*$) from X-ray [37] and nonlinear optical [51] measurements (not shown). (b) Definition of magnetic moment components. $\phi$ is defined as the angle between the $c$-axis and the total magnetic moment ($\boldsymbol{m}$).

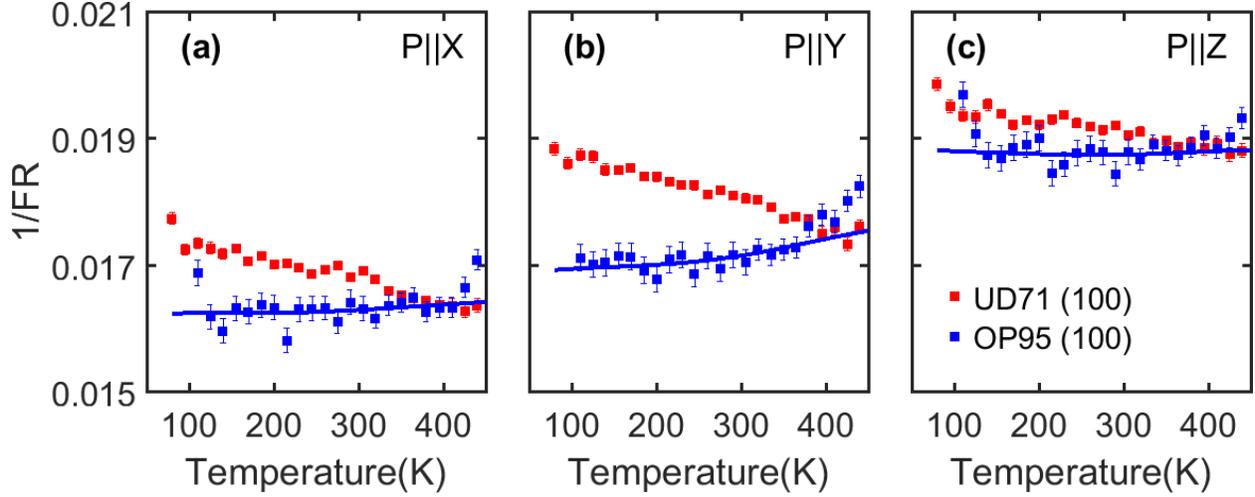

**Figure 2.** (a)-(c) Temperature dependence of the inverse flipping ratio (1/FR) at the (1 0 0) reflection for the three polarization directions for UD71 (red) and OP95 (blue). A magnetic signal is evident in UD71 from the upturn below $T_{q=0}$ = 360 - 380 K. For better visualization, the OP95 results are shifted by ~ -0.004 (X) -0.005 (Y) and ~-0.003 (Z) to best match the average of the UD71 data between 360 and 400 K (above $T_{q=0}$). Solid blue lines are smooth polynomial fits to the OP95 data, with less weight given to the high-temperature data (~ 400 K and higher), where the uncertainty in the flipping ratio increases due to thermal effects on the sample mount that turned out to be larger than for the measurement of UD71.

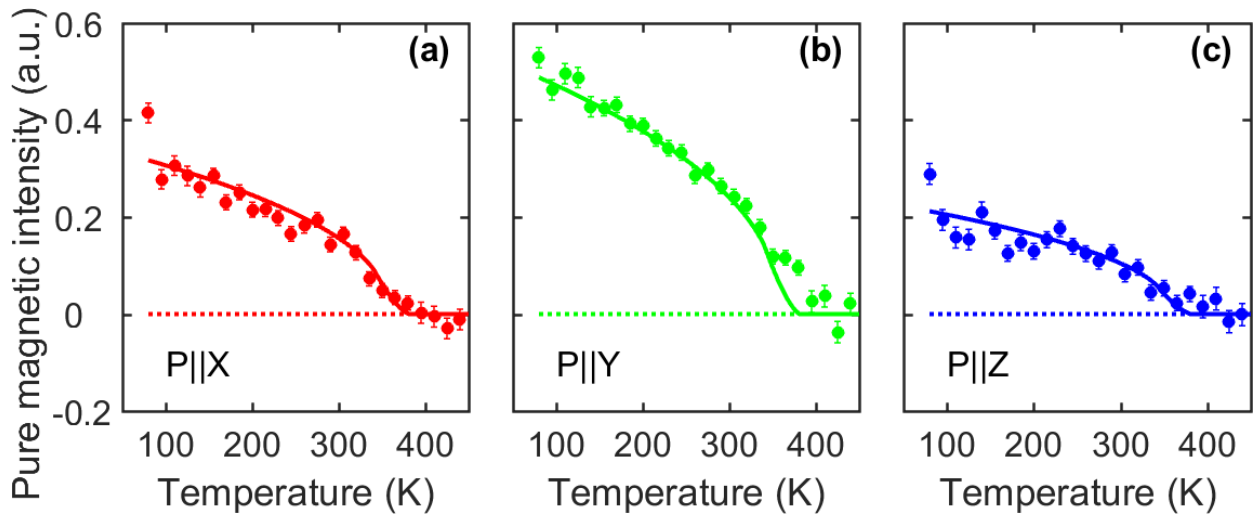

**Figure 3** (a)-(c) Temperature dependence of the (1 0 0) magnetic signal for UD71 for the three polarization directions. The signal is extracted according to Method 2, which simply assumes that no discernible magnetic Bragg signal exists in OP95. The fit results are shown as solid lines.

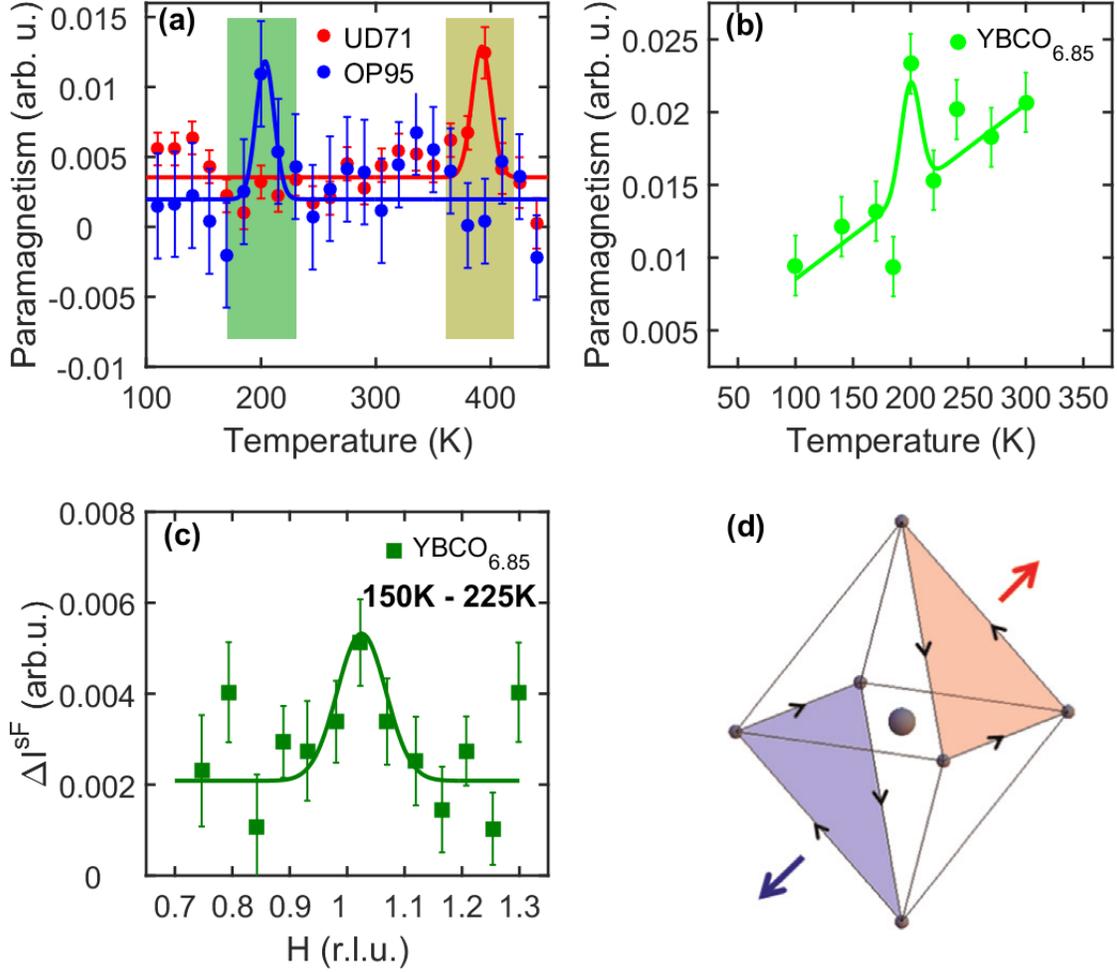

**Figure 4** (a) Temperature dependence of $\square_{SF} = I^x_{SF} + I^y_{SF} + I^z_{SF}$, the sum of the SF neutron cross-sections for the three polarization geometries. For both UD71 (red) and OP95 (blue), data are averaged over eleven momentum transfer values centered at **Q** = (0.88 0 -0.11) for better statistics. The peaks at ~ 370 K for UD71 and ~ 200 K for OP95, as highlighted by the yellow and green shaded areas, indicate the appearance of a magnetic signal away from the Bragg peak at a temperature consistent with the pseudogap temperature *T*\* obtained from transport measurements [38, 50] (see Supplementary Information for data analysis details). (b) Temperature dependence of $\square_{SF}$ for YBCO$_{6.85}$ measured with the same spectrometer with a similar analysis method (from [11]). The data show a peak at ~ *T*\* as well. (c) Difference, $\Delta\square^{SF}$, of **Q** = (*H* 0 -0.4) momentum scans for OP95 between *T* = 150 K and *T* = 225 K (see raw data in supplementary Fig. S7b). (d) Schematic of non-planar LC order with currents along the faces of oxygen octahedra consistent with the data for Hg1201 [23,46].

| Fitting Data | $m_c^2$ (a.u.) | $m_{ab}^2$ (a.u.) | $\phi = arctan\left(\frac{m_{ab}}{m_c}\right)$ |
|---|---|---|---|
| UD71 Method 1 | 0.20 ± 0.06 | 0.58 ± 0.14 | 71° ± 10° |
| UD71 Method 2 | 0.24 ± 0.09 | 0.68 ± 0.16 | 69° ± 10° |
| OP95 Method 1 | 0.06 ± 0.08 | 0.04 ± 0.04 | - |

**Table 1.** Fit results for the square of the in-plane and out-of-plane components of the magnetic moment determined (in the same arbitrary units) from polarization analysis involving the (1 0 0) reflection. For UD71, the results for both analysis methods are shown, including the angle of the moment direction with respect to [0 0 1].

| Compound [reference] | Sample $T_c$ | Estimated tilt angle | L |
|---|---|---|---|
| YBa$_2$Cu$_3$O$_{6.6}$  [6] | 61 K | 35 ± 7° | 0 |
|  |  | 55 ± 7° | 1 |
| YBa$_2$Cu$_3$O$_{6+x}$  [4] | 54 K, 61 K, 64 K | 45 ± 20° | 1 |
| YBa$_2$Cu$_3$O$_{6.85}$ [12] | 89 K | 40 ± 9° | 0.25 |
| La$_{1.915}$Sr$_{0.085}$CuO$_4$ [9] | 22 K | ~ 45° | 0 |
| HgBa$_2$CuO$_{4+\delta}$  [8] | 75 K | 45 ± 25° | 1 |
| Bi$_2$Sr$_2$CaCu$_2$O$_{8+\delta}$ [11] | 85 K | 20 ± 20° | 3 |

**Table 2.** Summary of previous estimates of the tilt angle of the magnetic moment for various underdoped cuprates based on measurements at for **Q** = (1 0 *L*) [3, 5, 8-12]. The estimated tilt angles all fall into the 45 ± 20° range.

**References**

[1] B. Keimer, S. A. Kivelson, M. R. Norman, S. Uchida and J. Zaanen, *Nature* **518**, 179 (2015)

[2] A. Kaminski, S. Rosenkranz, H. M. Fretwell, J. C. Campuzano, Z. Li, H. Raffy, W. G. Cullen, H. You, C. G. Oison, C. M. Varma, and H. Höchst, *Nature (London)* **416**, 610 (2002)

[3] B. Fauqué, Y. Sidis, V. Hinkov, S. Pailhès, C. T. Lin, X. Chaud, and P. Bourges, *Phys. Rev. Lett.* **96**, 197001 (2006)

[4] Y. Li, V. Balédent, N. Barišić, Y. Cho, B. Fauqué, Y. Sidis, G. Yu, X. Zhao, P. Bourges, and M. Greven, *Nature (London)* **455**, 372 (2008)

[5] H. A. Mook, Y. Sidis, B. Fauqué, V. Balédent, and P. Bourges, *Phys. Rev. B* **78**, 020506 (2008)

[6] V. Balédent, B. Fauqué, Y. Sidis, N. B. Christensen, S. Pailhès, K. Conder, E. Pomjakushina, J. Mesot, and P. Bourges. *Phys. Rev. Lett.* **105**, 027004 (2010)

[7] Y. Li, V. Balédent, N. Barišić, Y. C. Cho, Y. Sidis, G. Yu, X. Zhao, P. Bourges, and M. Greven, *Phys. Rev. B* **84**, 224508 (2011)

[8] P. Bourges and Y. Sidis, *C. R. Physique* **12**, 461 (2011)

[9] L. Mangin-Thro, Y. Sidis, P. Bourges, S. De Almeida-Didry, F. Giovannelli, and I. Laffez-Monot, *Phys. Rev. B* **89**, 094523 (2014)

[10] L. Mangin-Thro, Y. Sidis, A. Wildes and P. Bourges, *Nat. Commun.* **6**, 7705 (2015)

[11] L. Mangin-Thro, Yuan Li, Y. Sidis, and P. Bourges, *Phys. Rev. Lett.* **118**, 097003 (2017)

[12] Ph. Bourges, Y. Sidis, and L. Mangin-Thro, preprint, arXiv:1710.08173v3, to appear in Phys. Rev. B

[13] J. Xia, E. Schemm, G. Deutscher, S. A. Kivelson, D. A. Bonn, W. N. Hardy, R. Liang, W. Siemons, G. Koster, M. M. Fejer, and A. Kapitulnik, *Phys. Rev. Lett.* **100**, 127002 (2008)

[14] A. Shekhter, B. J. Ramshaw, R. X. Liang, W. H. Hardy, D. A. Bonn, F. F. Balakirev, R. D. McDonald, J. B. Betts, S. C. Riggs and A. Migliori *Nature* **498**, 75-77 (2013)

[15] Y. Lubashevsky, L. Pan, T. Kirzhner, G. Koren, and N.P. Armitage, *Phys. Rev. Lett.* **112**, 147001 (2014).

[16] L. Zhao, C. A. Belvin, R. Liang, D. A. Bonn, W. N. Hardy, N. P. Armitage and D. Hsieh. *Nat. Phys.* **13**, 250-254 (2017)

[17] H. Murayama, Y. Sato, R. Kurihara, S. Kasahara, Y. Mizukami, Y. Kasahara, H. Uchiyama, A.Yamamoto, E.-G.Moon, J. Cai, J. Freyermuth, M. Greven, T. Shibauchi, and Y. Matsuda, arXiv:1805.00276

[18] Jian Zhang, Z. F. Ding, C. Tan, K. Huang, O. O. Bernal, P.-C. Ho, G. D. Morris, A. D. Hillier, P. K. Biswas, S. P. Cottrell, H. Xiang, X. Yao, D. E. MacLaughlin, Lei Shu, *Sci. Adv*., **4**, 5235 (2018)

[19] C. M. Varma, *Phys. Rev. B* **55**, 14554 (1997)

[20] C. M. Varma, *Phys. Rev. B* **73**, 155113 (2006)

[21] C. Weber, T. Giamarchi, and C. M. Varma, *Phys. Rev. Lett.* **112**, 117001 (2014)

[22] Y. He and C.M. Varma, *Phys. Rev. B* **86**, 035124 (2012)

[23] C. Weber, A. Läuchli, F. Mila, and T. Giamarchi, *Phys. Rev. Lett.* **102**, 017005 (2009)

[24] S. Lederer and S. A. Kivelson, *Phys. Rev. B* **85**, 155130 (2012)

[25] V. M. Yakovenko, *Physica B* **460**, 159-164 (2015)

[26] S. W. Lovesey, D. D. Khalyavin and U. Staub, *J. Phys.: Condens. Matter* **27**, 292201 (2015)

[27] M. Fechner, M. J. A. Fierz, F. Thöle, U. Staub and N. A. Spaldin, *Phys. Rev. B* **93**, 174419 (2016)

[28] S. W. Lovesey and D. D. Khalyavin, *J. Phys.: Condens. Matter* **29**, 215603 (2017)

[29] C. M. Varma, *J. Phys.: Condens. Matter*, **26**, 505701 (2014).

[30] S. Chatterjee and S. Sachdev, *Phys. Rev. B* **95**, 205133 (2017).

[31] D. F. Agterberg, Drew S. Melchert, and M. K. Kashyap, *Phys. Rev. B* **91**, 054502 (2015).

[32] C. Morice, D. Chakraborty, X. Montiel, and C. Pépin, preprint, arXiv:1707.08497.

[33] J. W. Alldredge, K. Fujita, H. Eisaki, S. Uchida, and Kyle McElroy, *Phys. Rev. B* **87**, 104520 (2013).

[34] J.W. Loram, K.A. Mirza, J.R. Cooper, and W.Y. Liang, *Phys Rev Lett* **71**, 1740 (1993).

[35] D. Pelc, P. Popčević, G. Yu, M. Požek, M. Greven, N. Barišić, preprint, https://arxiv.org/abs/1710.10221

[36] X. Zhao, G. Yu, Y. Cho, G. Chabot-Couture, N. Barišić, P. Bourges, N. Kaneko, Y. Li, L. Lu, E. M. Motoyama, O. P. Vajk and M. Greven, *Adv. Mater.* **18**, 3243 (2006)

[37] W. Tabis, B. Yu, I. Bialo, M. Bluschke, T. Kolodziej, A. Kozlowski, E. Blackburn, K. Sen, E. M. Forgan, M. Zimmermann, Y. Tang, E. Weschke, B. Vignolle, M. Hepting, H. Gretarsson,

R. Sutarto, F. He, M. Le Tacon, N. Barišić G. Yu, and M. Greven, *Phys. Rev. B* **96**, 134510 (2017)

[38] N. Barišić, Y. Li, X. Zhao, Y. C. Cho, G. Chabot-Couture, G. Yu, and M. Greven, *Phys. Rev. B* **78**, 054518 (2008)

[39] N. Barišić, S. Badoux, M. K. Chan, C. Dorow, W. Tabis, B. Vignolle, G. Yu, J. Béard, X. Zhao, C. Proust and M. Greven, *Nat. Phys.* **9**, 761 (2013)

[40] M. K. Chan, M. J. Veit, C. J. Dorow, Y. Ge, Y. Li, W. Tabis, Y. Tang, X. Zhao, N. Barišić and M. Greven, *Phys. Rev. Lett.* **113**, 177005 (2014)

[41] E. van Heumen, R. Lortz, A. B. Kuzmenko, F. Carbone, D. van der Marel, X. Zhao, G. Yu, Y. Cho, N. Barišić, M. Greven et al., Phys. Rev. B 75, 054522 (2007)

[42] M. K. Chan, C. J. Dorow, L. Mangin-Thro, Y. Tang, Y. Ge, M. J. Veit, G. Yu, X. Zhao, A. D. Christianson, J. T. Park, Y. Sidis, P. Steffens, D. L. Abernathy, P. Bourges and M. Greven, *Nat. Commun.* **7**, 10819 (2016)

[43] M. K. Chan, Y. Tang, C. J. Dorow, J. Jeong, L. Mangin-Thro, M. J. Veit, Y. Ge, D. L. Abernathy, Y. Sidis, P. Bourges and M. Greven, *Phys. Rev. Lett.* **117**, 277002 (2016)

[44] A. Pal, S.R. Dunsiger, K. Akintola, A. Fang, A. Elhosary, M. Ishikado, H. Eisaki, J. E. Sonier, Phys. Rev. B 97, 060502 (2018)

[45] S. Di Matteo and M. Norman, *Phys. Rev. B* **85**, 253143 (2012)

[46] V. L. Aksenov, A. M. Balagurov, V. V. Sikolenko, V. G. Simkin, V. A. Alyoshin, E. V. Antipov, A. A. Gippius, D. A. Mikhailova, S. N. Putilin, and F. Bouree, *Phys. Rev. B* **55**, 3966 (1997)

[47] A. M. Balagurov, D. V. Sheptyakov, V. L. Aksenov, E. V. Antipov, S. N. Putilin, P. G. Radaelli, and M. Marezio, *Phys. Rev. B* **59**, 7209 (1999)

[48] J. Orenstein, *Phys. Rev. Lett.* **107** 067002 (2011)

[49] A. Yamamoto, W.-Z. Hu, and S. Tajima, *Phys. Rev. B* **63**, 024504 (2000)

[50] N. Barišić, M. K. Chan, Y. Li, G. Yu, X. Zhao, M. Dressel, A. Smontara and M. Greven, Proc. Natl. Acad. Sci. **110** 12235 (2013)

[51] J. P. Hinton, E. Thewalt, Z. Alpichshev, F. Mahmood, J. D. Koralek, M. K. Chan, M. J. Veit, C. J. Dorow, N. Barišić, A. F. Kemper, D. A. Bonn, W. N. Hardy, Ruixing Liang, N. Gedik, M. Greven, A. Lanzara, and J. Orenstein, *Sci. Rep.* **6**, 23610 (2016)

Supplemental Material for

# Orientation of the intra-unit-cell magnetic moment in the high-$T_c$ superconductor $HgBa_2CuO_{4+\delta}$

Yang Tang[1], Lucile Mangin-Thro[2], Andrew Wildes[2], Mun K. Chan[1], Chelsey J. Dorow[1], Jaehong Jeong[3], Yvan Sidis[3], Martin Greven[1], Philippe Bourges[3]

[1] *School of Physics and Astronomy, University of Minnesota, Minneapolis, MN 55455, USA*

[2] *Institut Laue-Langevin, 71 avenue des martyrs, Grenoble 38000, France*

[3] *Laboratoire Léon Brillouin, CEA-CNRS, Université Paris-Saclay CEA-Saclay, Gif sur Yvette 91191, France*

**Content:**

1. Sample description and characterization
2. Experimental information and data analysis methods of the D7 (ILL) measurements
3. Magnetic scattering at position away from the Bragg spots, Q=(0.88 0 -0.11)
4. Additional measurements on the triple-axis 4F1 (LLB/Orphée)
5. Supplementary Figures S1-7
6. References

## 1. Sample description and characterization

Each of the two samples is comprised of approximately 30 co-aligned single crystals (each sample with a total mass of about 2 g). The crystals were grown by a flux method [38] and subsequently subjected to a heat treatment [33] in order to achieve the desired $T_c$. The superconducting transition temperature for each sample was estimated by averaging magnetic susceptibility data of individual crystals measured by Quantum Design, Inc., Magnetic Property Measurement System (MPMS). The result of this averaging is shown in Fig. S1. We estimate $T_c$ = 71 ± 3 K (sample labeled UD71) and $T_c$ = 95 ± 3 K (sample labeled OP95). At the (1 1 0) Bragg reflection, we determined full-width-at-half-maximum (FWHM) mosaics of 1.5° for UD71 and 2.5° for OP95. Measurements were carried out at temperatures that ranged from slightly above $T_c$ up to about 450 K, the temperature up to which the glue used to mount crystals (GE varnish) was found to be stable.

## 2. Experimental information and data analysis methods of the D7 measurements

We describe here the experimental method to perform the data analysis to extract the $\mathbf{q} = 0$ magnetic signal. In contrast with previous reports for $HgBa_2CuO_{4+\delta}$ (Hg1201) [1,2], we used a different spin-polarized neutron diffractometer that enabled systematic polarization analysis [3,4] at every temperature: the cold-neutron diffractometer D7 at the Institute Laue-Langevin (ILL), Grenoble, France. The set-up of the experiment was similar to that of a previous study of $YBa_2Cu_3O_{6+x}$ [5] and is shown in Fig. S2a. However, in order to minimize neutron absorption of Hg, the incident neutron beam was monochromated to a relatively long wavelength (incident neutron energy of $E_i$ = 20 meV, incident wavelength 3.1 Å or wave vector ≈ 2.02 Å$^{-1}$). This had the effect of a broader momentum resolution compared to the previous report [5,6] where the incident neutron wavelength was 4.8 Å. We quote the scattering wave-vector $\mathbf{Q} = H\mathbf{a}^* + K\mathbf{b}^* + L\mathbf{c}^* \equiv (H\ K\ L)$ in reciprocal lattice units, where $\mathbf{a}^* = \mathbf{b}^* = 1.62$ Å$^{-1}$ and $\mathbf{c}^* = 0.66$ Å$^{-1}$ are the room-temperature values related to the lattice parameters of the $HgBa_2CuO_{4+\delta}$ system. The samples were mounted such that Bragg peaks ($H\ 0\ L$) were accessible.

D7 has fixed polarization directions for the neutron beam along the X-, Y- and Z-directions of a Cartesian coordinate system, where Z is perpendicular to the scattering plane (see Fig. S2.a). The initial state of the neutrons was prepared with a supermirror polarizer, a Mezei flipper, and subsequent small guide fields generated by Helmholtz coils to adiabatically rotate the spin to the desired polarization direction. The Helmholtz coil system differed from the one used previously [5], as in our case the guide field within the scattering plane was produced with four coils; as a result, the magnetic guide field generated at the sample position was more homogeneous than in the prior work.

Detailed measurements of the temperature dependence of a few Bragg peaks were performed along the three neutron polarization directions (X, Y, and Z), both in the spin-flip (SF) and the non-spin-flip (NSF) channel. The $\mathbf{q} = 0$ magnetic signal is expected to be present at the (1 0 $L$) reflections [5]. In principle, magnetic scattering occurs in both SF and NSF channels. However, in our case of relatively weak magnetic intensities, nuclear Bragg scattering dominates in the NSF channel. Therefore, magnetic scattering is only observed in the SF channel. Ideally, nuclear scattering only contributes to the measured intensity in the NSF channel. However, due to experimental limitations, the neutron beam is not perfectly polarized and analysed, which results in a "leakage" of NSF neutrons into the SF channel. A measurement of this effect is the flipping ratio ($FR$), which is defined as the ratio between the NSF and SF intensities measured at a reference nuclear Bragg peak. The available 66 supermirror benders used to analyse the scattered neutron beam do not have identical polarization efficiency. We chose the best bender that matched the $Q$-range of the Bragg peaks in order to achieve the best possible flipping ratio ($FR \sim 40$). Moreover, we used two of the triple-blade pyrolytic graphite monochromators that D7 is equipped with for better polarization efficiency and stability.

We studied two Hg1201 samples, one underdoped (UD71) and one nearly optimally doped (OP95). For both samples, we measured at the (1 0 0) and (0 0 3) Bragg peaks, as well as at the (2 0 0) reflections. The measurements at (1 0 0) and (0 0 3) can be readily compared, as the two reflections have similar values of $Q$, which allowed the use of the same bender-detector couple to obtain the highest $FR$. In addition, we measured a plate of pyrolytic graphite (PG002) to characterize the thermal dependence of $FR$ in a standard non-magnetic sample in a scattering geometry similar to that used to measure the Hg1201 samples.

In Figs. S3 and S4, we report raw data obtained on both Hg1201 samples as well on the PG002 sample for selected Bragg reflections. In all cases, the nuclear NSF intensity was maximized at 100 K in the detector with the highest flipping ratio. Both the sample rotation abngle and the detector bank were scanned to maximize the NSF intensity at 100 K. In Fig. S3, the NSF intensity is shown versus temperature for the three polarizations X, Y and Z, and no difference between polarization is observed. However, the nuclear Bragg intensity does not simply decrease with temperature, as would be expected from the Debye-Waller factor.

In order to analyse our data and describe this behavior, we write the temperature dependence of the SF and NSF intensities as:

$$I^{NSF}_{X,Y,Z}(T) = I^{NSF}_{X,Y,Z}(T=0)f(T) \tag{1}$$

$$I^{SF}_{X,Y,Z}(T) = [BG^{SF}_{X,Y,Z} + M_{x,y,z}(T)]f(T) \tag{2}$$

where $f(T)$ is a thermal envelope function that captures contributions to the measured intensity due to thermal effects such as the relative motion of the sample with regard to the incident beam (and hence the projection of the scattered beam onto the detector) due to thermal contraction/expansion of the sample stick, as well as the Debye-Waller factor. Also, the thermal dependence of the lattice parameters is not fully captured at all temperatures by the momentum resolution. With these combined thermal effects captured by this envelope function, the NSF intensity then becomes independent of temperature and equal to the intrinsic value $I^{NSF}_{X,Y,Z}(T=0)$ given by Bragg scattering at $T=0$. Note that the NSF intensities do contain a magnetic contribution, but this can be neglected as it is very weak with respect to the nuclear contribution. As shown in Fig. S3, the function $f(T)$ (Eq. (1)) describes the average of the NSF data obtained in the three polarization geometries. The envelope function $f(T)$ and the intrinsic Bragg scattering intensity $I^{NSF}_{X,Y,Z}(T=0)$ are estimated via polynomial fits to the NSF data. A single function $f(T)$, normalized to 1 at low temperature, provides an excellent fit for all three polarizations. Figure S3 shows $I^{NSF}_{X,Y,Z}(T)$ for both UD71 and OP95 as well as for PG002. A different envelope function, $f(T)$, is nevertheless obtained for each Bragg reflection and sample due to the combined thermal effects discussed above.

Next, the thermal envelope function, $f(T)$, defined for the NSF geometry is used for the SF intensities following Eq. (2). We scale the SF data by $f(T)$ along X, Y and Z, as shown in

Fig. S4. The estimation of $f(T)$ and the scaling of the SF intensities is done separately for each sample and Bragg peak. Figure S4 shows the normalized SF intensities, $\frac{I_{X,Y,Z}^{SF}(T)}{f(T)}$, for each polarization. For each polarization geometry, the normalized SF response consists of two components (see Eq. (2)): a background term $BG_{X,Y,Z}^{SF}$ that results from the unavoidable leak of NSF scattering into the nominal SF channel, and the genuine magnetic intensities $M_{X,Y,Z}(T)$. The background term can be written as:

$$BG_{x,y,z}^{SF} = \frac{I_{X,Y,Z}^{NSF}(T=0)}{FR_{X,Y,Z}(T)} = \frac{I_{X,Y,Z}^{NSF}(T=0)}{FR_{X,Y,Z}(T=0)} g(T) \qquad (3)$$

where $FR(T)$ is the temperature-dependent flipping ratio which, in principle, should be independent of the temperature. However, as shown in Fig. S2, $FR(T)$ exhibits temperature dependence even when no magnetic contribution is expected (see, *e.g.*, the (0 0 3) Bragg peak of Hg1201 and the (0 0 2) reflection for graphite). This is due to polarization inhomogeneities of the beam and the fact that the sample slightly moves with temperature relative to the neutron beam. However, the same thermal dependence is found for the the polarizations. It can then be represented by a single function $g(T)$ (see Eq. (3)) independent of the polarization. We note that $g(T)$ differs from $f(T)$ in the sense that, although both originate from imperfections of the instrument, the former arises during the spin-polarization and detection process, whereas the latter results from the thermal variation of both instrument position and sample lattice parameters. Note that the inverse flipping ratio $1/FR_{X,Y,Z} = I_{X,Y,Z}^{SF}/I_{X,Y,Z}^{NSF} = g(T)/FR_{X,Y,Z}(T=0) + M_{X,Y,Z}(T)/I_{X,Y,Z}^{NSF}(T=0)$, represented in Fig. 2 of the manuscript, is independent of *f(T)*, but depends on *g(T)*.

As shown in Fig. S4, $g(T)$ is typically described by a simple linear behaviour, $1 + \varepsilon T$; this describes the data for the (0 0 3) reflection of UD71 and for the (0 0 2) reflection of graphite. For OP95, the fit results show linear behavior as well, consistent with the absence of any discernable magnetic signal. However, the slope $\varepsilon$ differs from that for graphite or the (0 0 3) reflection of UD71. This is likely related to differences in sample and scattering geometry: due to the approximate square shape of the sample, the (0 0 *L*) Bragg peaks are in reflection whereas the (*H* 0 0) peaks are in transmission. Moreover, no magnetic signal is discernible in the measurements of a piece of graphite (see Fig. S4) used as a reference, which further demonstrates the stability of our measurements.

In contrast to all these data, the (1 0 0) refection for UD71 exhibits a different non-linear thermal evolution (Fig. S4), which indicates a clear magnetic signal with an onset temperature of about 360 K in all three polarization directions. In Eq. (2), such a magnetic term is obtained through the longitudinal polarization analysis of the multi-detector diffractometer [3-5]. It is related to the different contributions of the **q** = 0 magnetic moment ***m*** in the three polarization geometries. We first need to write the magnetic moment as a superposition of moments along the reciprocal lattice basis, $m^2 = m_a^2 + m_b^2 + m_c^2$. Since Hg1201 has tetragonal symmetry, and hence ***a**** and ***b**** are equivalent, we can simply express the moment in terms of in terms of its in-plane and out-of-plane components: $m^2 = m_{ab}^2 + m_c^2$ with $m_{ab}^2 = 2m_a^2$ (Fig. S1c). With α defined as the angle between momentum transfer ***Q*** and the polarization direction X (Fig. S2.b), the magnetic moment components along the three polarization directions are:

$$M_z \propto m_c^2 \qquad (4)$$

$$M_y \propto \frac{1}{2}m_{ab}^2 + \sin^2\alpha \, m_c^2 \qquad (5)$$

$$M_x \propto \frac{1}{2}m_{ab}^2 + \cos^2\alpha \, m_c^2 \qquad (6)$$

These relations are specific to only magnetic scattering in neutron diffraction. They respect that only the components of the magnetic moment perpendicular to ***Q*** are observed in neutron scattering. Figure S2.b shows that the angle between ***Q*** and X is $\alpha = \frac{\pi}{2} - \theta + \gamma$, where $\theta$ is the Bragg scattering angle and $\gamma$ is the angle between incident beam and X, which is fixed to be 41.6° on D7. Therefore, we obtain α = 108.2° ± 5° for ***Q*** = (1 0 0), where the uncertainty comes from both the sample mosaic ($\theta$) and the instrument ($\gamma$).

Prior polarized-neutron diffraction work demonstrated an order-parameter-like temperature dependence [2,5] for the **q** = 0 magnetic moment, and we therefore write:

$$m_{ab,c}(T) = m_{ab,c}\left(1 - \frac{T}{T_{q=0}}\right)^\beta \qquad (7)$$

The SF intensities in Eq. 2 can then be written as:

$$I_X^{SF}(T) = [\frac{I_X^{NSF}(T=0)}{FR_X(T=0)}g(T) + (\frac{1}{2}m_{ab}^2 + 0.0976\, m_c^2)\left(1 - \frac{T}{T_{q=0}}\right)^{2\beta}]f(T) \qquad (8)$$

$$I_Y^{SF}(T) = [\frac{I_Y^{NSF}(T=0)}{FR_Y(T=0)}g(T) + (\frac{1}{2}m_{ab}^2 + 0.9024\, m_c^2)\left(1 - \frac{T}{T_{q=0}}\right)^{2\beta}]f(T) \qquad (9)$$

$$I_Z^{SF}(T) = [\frac{I_Z^{NSF}(T=0)}{FR_Z(T=0)}g(T) + m_c^2\left(1 - \frac{T}{T_{q=0}}\right)^{2\beta}]f(T) \qquad (10)$$

One can now use Eqs. 8-10 to analyse our data of Fig. S3. Two distinct methods are actually used to analyse our scaled SF data:

- Method 1 is a "blind" test, which assumes that both UD71 and OP95 exhibit the same linear SF background temperature dependence: $g_{UD71}(T) = g_{OP95}(T)$, that means they share the same slope. A **q** = 0 magnetic signal is allowed in this analysis for both samples.
- Method 2 assumes that OP95 exhibits no signal or even that the ordered moment for OP95 is immeasurably small consistent with prior work [1]. We then use the OP95 data to analyze the UD71 result at the same Bragg peak (1 0 0).

In method 1, we simultaneously fit the data for each sample which involves 8 fitting parameters for the and for all 3 polarization directions: $FR_{X,Y,Z}(T=0)$, $\varepsilon$, $m_{ab}^2$, $m_c^2$, $T_{q=0}$ and $\beta$. The first four parameters describe completely the intrinsic beam polarization of the instrument whereas the last four parameters fully determine the magnetic intensity. The results shown in Fig. 2 of the manuscript reproduced the data shown in Fig. S3 fitted by method 1.

Analysis method 2 uses the OP95 data to analyze the UD71 result. In particular, given that the ordered moment for OP95 is immeasurably small (as confirmed by method 1 and consistent with prior work [1]), and the fact that the sample size and shape as well as the measurement geometry for UD71 and OP95 are very similar, we assume that the OP95 data serve as good reference of $g(T)$, and that the UD71 data feature magnetic signal on top of a background signal that has the same temperature dependence. In other words, the SF data for UD71 are still expressed by Eqs. (8) – (10), but for OP95 they collapse to Eq. (3). Note that although the two data sets are assumed to share the same $g(T)$, $BG_{X,Y,Z}^{SF}$ may differ due to different values of $I^{NSF}_{X,Y,Z}(T=0)$ as a result of different sample masses. With $I^{NSF}_{X,Y,Z}(T=0)$ estimated for OP95 (see Fig. S3) and Eq. 3, we determine $g(T)$. Next, the background for UD71 can be calculated using $g(T)$ and $I^{NSF}_{X,Y,Z}(T=0)$ as estimated for UD71. We then subtract the estimated background from the SF data scaled by the envelope function (Fig. S4 d-f), and obtain the pure magnetic signal shown in Fig. 3 of the manuscript. The data in all 3 polarization directions are fit simultaneously, with fit parameters $m_{ab}^2$, $m_c^2$, $T_{q=0}$ and $\beta$.

The superconducting transition width of the UD71 sample is ~ 4 K defined as the difference between the two temperatures at 10% and 90% of the maximum susceptibility magnitude, which corresponds to a variation of $T_{q=0}$ of about 30 K (see Fig. 1 of the manuscript) and should lead to a rounding of the pseudogap transition. Our fit in Fig. 3 of the manuscript takes this rounding into account assuming a Gaussian distribution of transition temperatures. For UD71, our overall estimate of the effective exponent $\beta \approx 0.225 \pm 0.075$ is consistent with the prior work for Hg1201 [2] and with the value $\beta \approx 0.185 \pm 0.060$ estimated for YBCO [7]. Interpreted as an order-parameter exponent, the value for Hg1201 is not inconsistent with typical 3D-critical exponents, which lie in the $\beta = 0.30$-$0.36$ range [8]. We note that the LC model has been argued to belong to the universality class of the two-dimensional Ashkin-Teller-model, for which order-parameter exponents in the range from 1/8 to 1/4 are possible [9, 10]. However, we emphasize that the data in Fig. 4 of the manuscript suggest that the critical regime is rather narrow, whereas the fit to Eq. (4) in Fig. 3 extends over a very wide temperature range, and thus $\beta$ should probably not be viewed as a critical exponent.

## 3. Magnetic scattering at position away from the Bragg spots, Q=(0.88 0 -0.11)

D7 is a multi-detector diffractometer that covers a wide scattering angle range [3-4]. So far, we have discussed the data obtained with detectors that correspond to the Bragg position (1 0 0). However, additional information can be extracted from detectors that correspond to $Q$ values away from the Bragg spots. In particular, at $\mathbf{Q} = (0.88\ 0\ -0.11)$, diffuse magnetic scattering can be observed. Similar analysis was performed for YBCO [5]. We reproduced in Figure S5a these data by plotting the sum of all SF cross-sections, $\Sigma_{SF} = I^X_{SF} + I^Y_{SF} + I^Z_{SF}$, summing intensities from four detectors near $\mathbf{Q} = (0.88\ 0\ -0.11)$. It is found that $\Sigma_{SF}$ exhibits a peak at the onset temperature of the $\mathbf{q} = 0$ magnetic order. This peak is indicative of magnetic critical slowing down at $T_{q=0}$ of the IUC magnetic order. For both UD71 and OP95 samples, the same quantity $\Sigma_{SF}$ for **Q** = (0.88 0 -0.11) also displays a peak at a temperature close to the pseudogap temperature $T^*$. In Fig. 4a, we report this signal, with background removed. The background is determined by the polarization analysis of the three SF cross sections according to Eqs. (4)-(6) with $\alpha = 105° \pm 5°$ for **Q** = (0.88 0 -0.11). Another way to reveal this feature is to compute the first derivative of $\Sigma_{SF}$, as shown in Fig. S5b: a sharp S-shape is seen for both samples, with a characteristic temperature that matches $T^*$ within error.

## 4. Additional measurements on the triple-axis 4F1

Additional measurements were performed on the triple-axis spectrometer 4F1 at LLB/Orphée. This instrument is equipped with polarized-neutron capabilities [1,2,5] and XYZ longitudinal polarization analysis. We recall that the angle $\alpha$ between ***Q*** and X (Fig. S2 and Eq. 4-6) goes to zero for a triple-axis spectrometer. The X channel thus corresponds to the **P**//**Q** configuration, and the full magnetic intensity is given by $M_X$. The polarization sum rule then reads: $M_X = M_Y + M_Z$.

Figure S6 shows the temperature dependence of the **q** = 0 magnetic intensity at **Q** = (1 0 L). Measurements were performed with **P**//**Q** in the SF channel (where the magnetic signal is expected to be maximum). In Figs S6a-b, a magnetic signal appears below $T_{q=0} \sim 220$ K. This signal displays a characteristic order-parameter-like $T$-dependence, Eq. 7 (with $\beta = 0.25$). Converted to absolute units, the magnetic intensity is estimated to be ~2 mbarns at **Q**=(1 0 0) and ~0.5mbarns at **Q**=(1 0 1), in agreement with our upper bound from the D7 measurement. Systematic errors in the amplitude larger than the statistical errors are possible due to remaining uncertainties in the determination of the reference of the bare flipping ratio (Fig. S6).

The Fig, S7b shows a momentum scan in the SF channel for **P**//**Q** along (H 0 -0.4) at two temperatures, 150 K and 225 K. The weak peak maximum at $H = 1$ at 150 K indicates short-range **q** = 0 magnetic order. Similar momentum scans have been used as well in $La_{1.9}Sr_{0.1}CuO_4$ to reveal short-range IUC magnetic order [11].

Figure S7a shows a polarization analysis in the SF channel at **Q** = (0.9 0 0) for OP95. The three polarizations XYZ were measured on the triple-axis spectrometer (where $\alpha = 0$ in Eqs. (4)-(6)). The quantity shown is $\Delta_{SF} = 2I_{SF}^x - (I_{SF}^y + I_{SF}^z)$, which is interestingly fully independent of background contributions. Therefore, $\Delta_{SF} = \frac{1}{2}<m_{ab}^2> + <m_c^2>$, where the brackets imply the time average of the square of the IUC fluctuations. The peak in temperature indicates the emergence of IUC critical fluctuations around ~ 220 K, in good agreement with the D7 measurement, despite the limited counting statistics.

## 5. Supplementary Figures

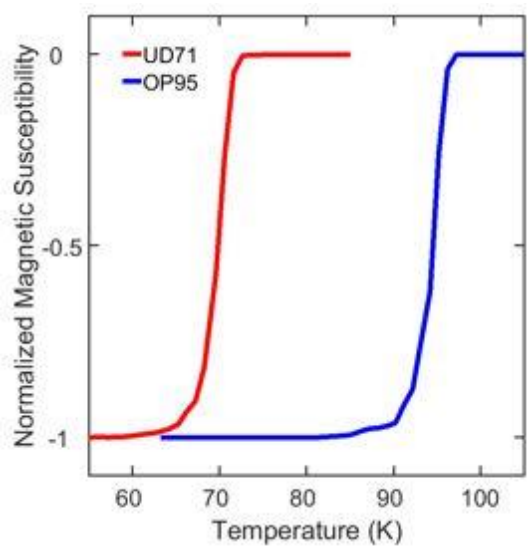

**Figure S1.** Magnetic susceptibility for the two samples studied in the present work, each comprised of about 30 co-aligned crystals. Data were obtained for individual crystals with a *c*–axis magnetic field of 5 G, averaged (weighted by crystal mass), and the result was then normalized to -1. The transition temperatures for the two samples labeled UD71 and OP95 are estimated from the transition midpoints, which are $T_c = 71$ K and $T_c = 95$ K, respectively.

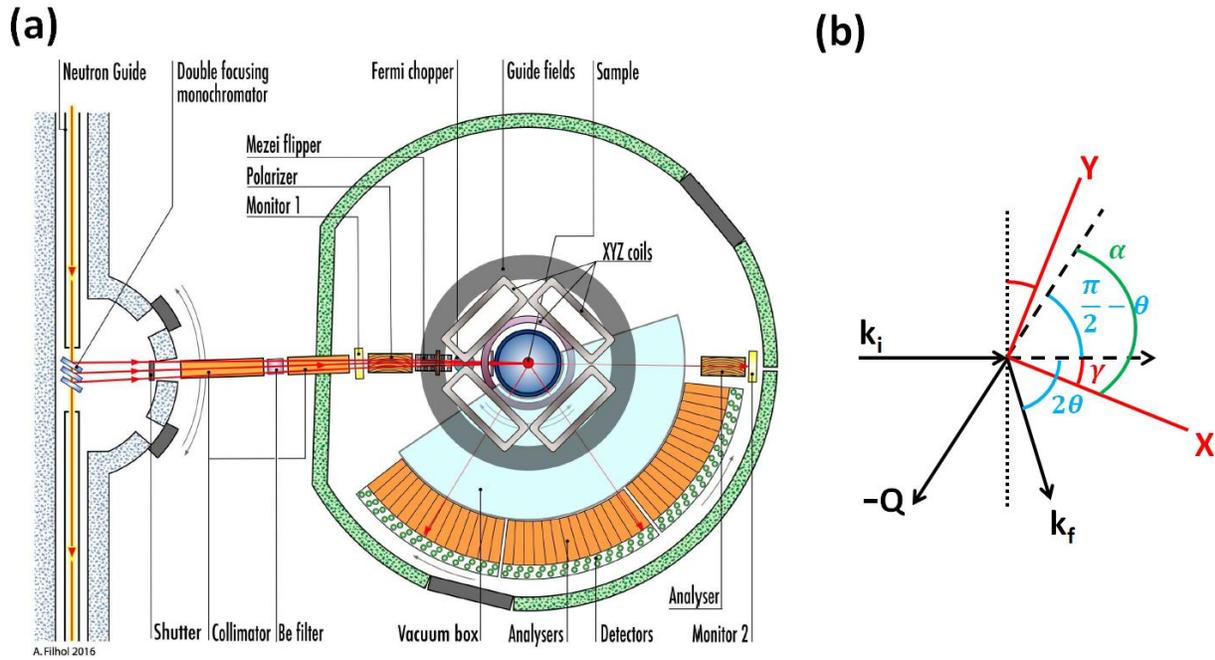

**Figure S2.** (a) Schematic of the D7 diffractometer at ILL (from ref. 6). (b) Schematic of scattering plane, defined by the incident ($k_i$) and scattered ($k_f$) neutron wave vectors, along with the definition of the polarization directions X and Y (Z is perpendicular to the scattering plane). The angle $\gamma = 41.6°$ is set by default instrument configuration, $2\theta$ is the scattering angle, and $\alpha$ is defined as the acute angle between the momentum transfer $\mathbf{Q} = \mathbf{k_i} - \mathbf{k_f}$ and the polarization direction X.

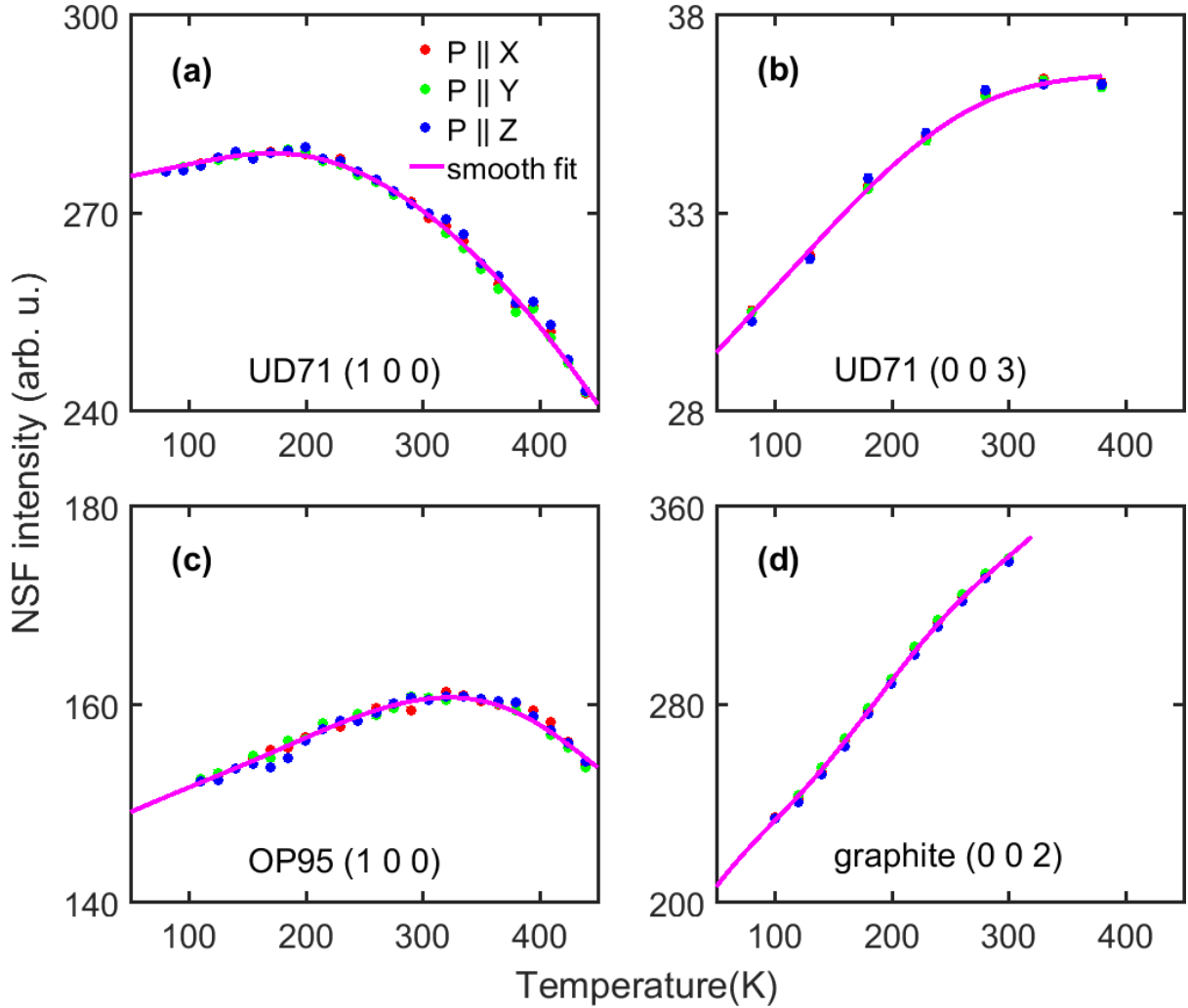

**Figure S3.** Temperature dependence of the NSF data for the four Bragg peaks/samples measured in this work. (a)-(d) UD71 at (1 0 0); UD 71 at (0 0 3); OP95 at (1 0 0) and graphite at (0 0 2), respectively, as indicated by the label at bottom of each figure. Red, green and blue circles pertain to incident neutron spin parallel to X, Y and Z directions. Magenta solid lines: envelope function $f(T)$ obtained from fits to the scaled average of the NSF data. No data were obtained at temperatures below $T_c$. At lower temperatures, the lines are extrapolations from fits to the data above $T_c$.

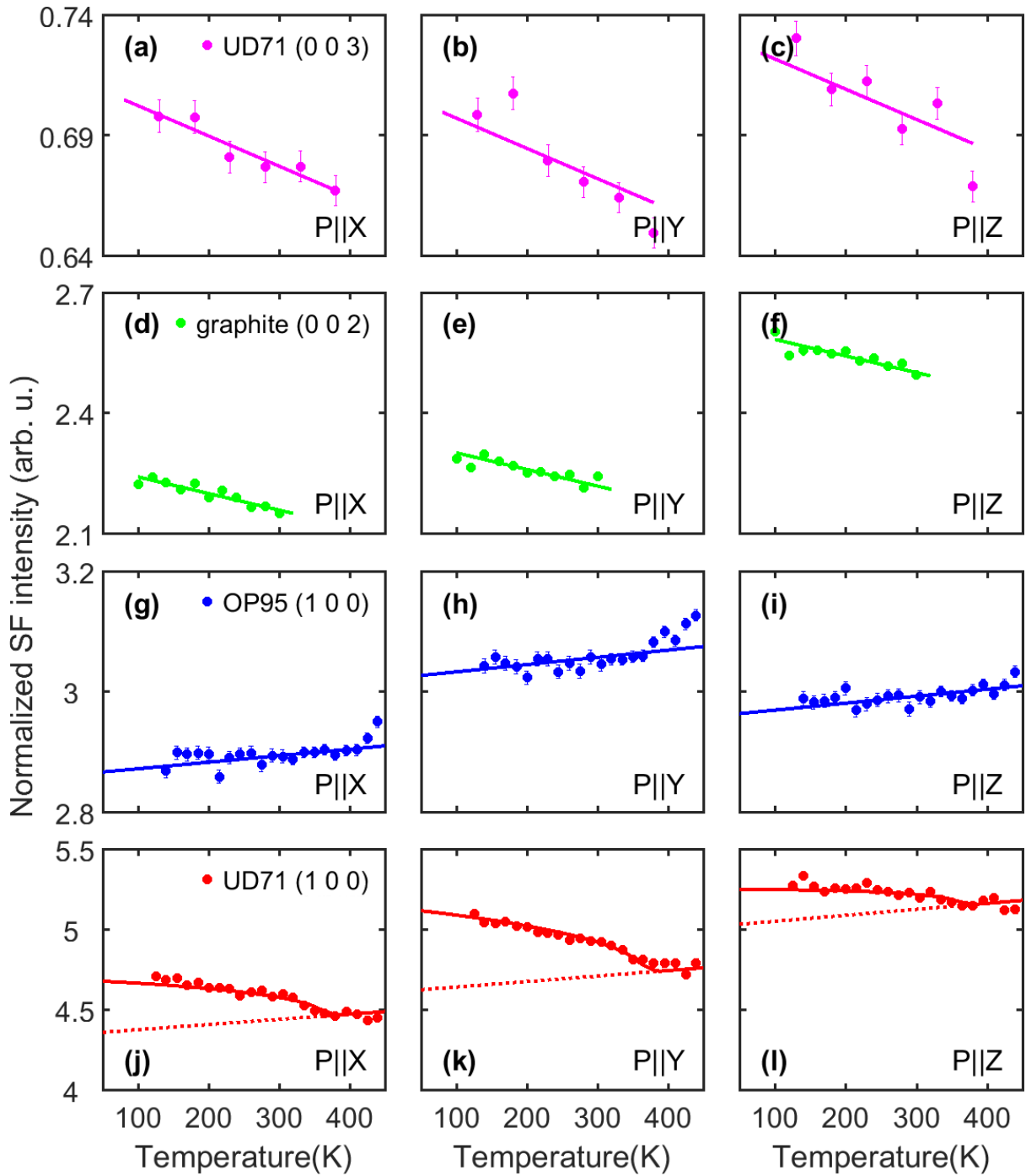

**Figure S4.** Temperature dependence of the scaled SF data for the Bragg peaks/samples measured in this work. Rows from top to bottom: UD71 at (0 0 3), graphite at (0 0 2), OP95 at (1 0 0), and UD71 at (1 0 0). Columns from left to right: incident neutron spin polarized parallel to the X, Y and Z directions. Solid lines for UD71 at (0 0 3) and graphite at (0 0 2) are linear fits of the data. Solid lines for OP95 and UD71 at (1 0 0) are fits using method 1 (blind test) as described in the main text. The red dashed lines show the linear background for the fits, which are fixed to be the same with those for OP95 results by assumption.

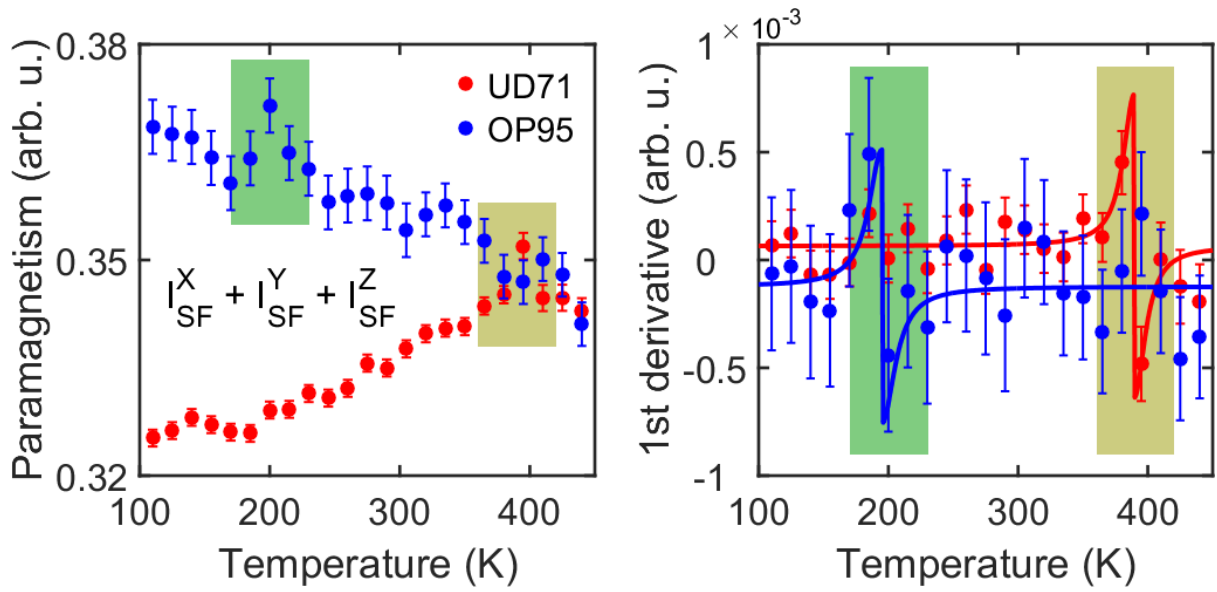

**Figure S5.** (a) Sum of the three cross sections. (b) First derivate of the sum of the signal in (a). The change of sign better illustrates the characteristic magnetic critical slowing down at $T_{q=0}$.

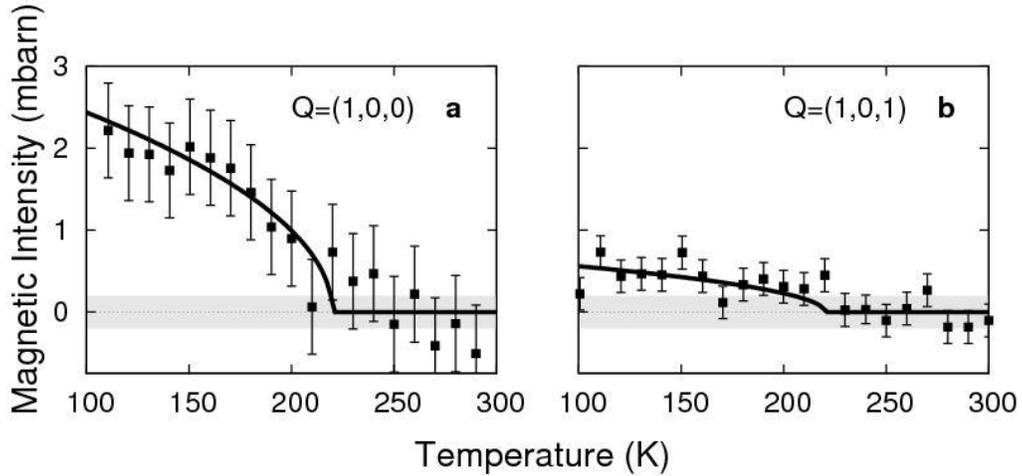

**Figure S6** (a): Temperature dependence of the **q**=0 magnetic intensity extracted from measurements at the Bragg reflections **Q**=(1 0 *L*): (a) *L* = 0 and (b) *L* = 1. All the measurements were carried out with the **P**//**Q** configuration in the SF channel on the spectrometer 4F1 at LLB/Orphée on a OP95 sample that was very similar to the one measured on the D7 instrument at ILL. The data were obtained using a bare flipping ratio reference, obtained at (2 0 0) and (0 0 4) where no magnetic signal is expected, and further averaged following ref. [5] for YBCO. The magnetic intensity at (1 0 *L*) is calibrated in absolute units using the intensity of the nuclear Bragg peaks (4.62 barns at *L* = 0 and 1.26 barns at *L* = 1 [1,2]). Error bars are of standard deviation.

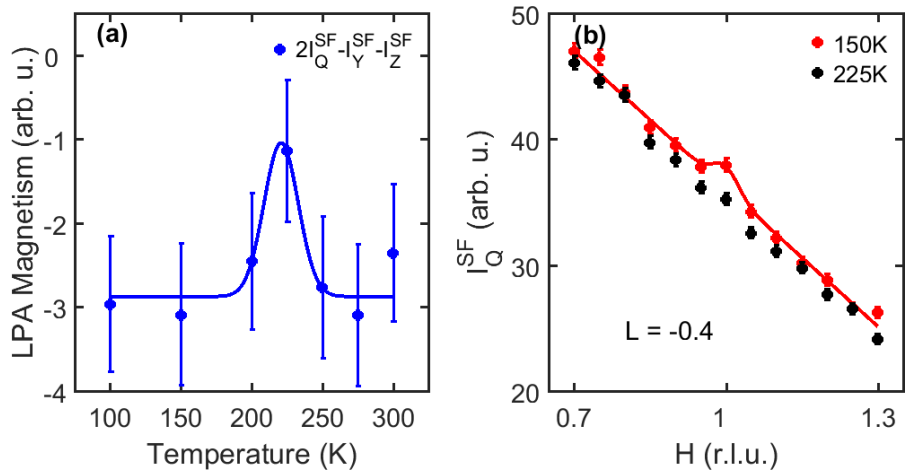

**Figure S7.** (a) Temperature dependence of longitudinal polarization analysis (LPA) of pure magnetic signal measured on spectrometer 4F1 at LLB for the same OP95 sample used on D7, at a momentum transfer $\mathbf{Q} = (0.9\ 0\ 0)$, off the Bragg peak. (b) Momentum scan across (1 0 -0.4) with incident neutron spin polarized along the momentum transfer $\mathbf{Q}$ ($\mathbf{H}//\mathbf{Q}$). Red and black data are measured at 150 K and 225 K, respectively.

## 5. References

[1] Y. Li, V. Balédent, N. Barišić, Y. Cho, B. Fauqué, Y. Sidis, G. Yu, X. Zhao, P. Bourges, and M. Greven, *Nature (London)* **455**, 372 (2008)

[2] Y. Li, V. Balédent, N. Barišić, Y. C. Cho, Y. Sidis, G. Yu, X. Zhao, P. Bourges, and M. Greven, *Phys. Rev. B* **84**, 224508 (2011)

[3] Stewart, J.R. *et al.* Disordered materials studied using neutron polarization analysis on the multi-detector spectrometer, D7. *J. Appl. Cryst.* 42, 69-84 (2009).

[4] G. Ehlers, G. Stewart, J.R., Wildes, A.R., Deen, P.P. & Andersen, K.H. Generalization of the classical xyz-polarization analysis technique to out-of-plane and inelastic scattering. *Review of Scientific Instruments* 84, 093901 (2013).

[5] L. Mangin-Thro, Y. Sidis, A.R. Wildes and P. Bourges, *Nat. Commun.* **6**, 7705 (2015)

[6] T. Fennell, L. Mangin-Thro, H. Mutka, G.J. Nilsen and A.R. Wildes, Nucl. Inst. and Meth. Phys. Res. A **857** 24 (2017)

[7] H. A. Mook, Y. Sidis, B. Fauqué, V. Balédent, and P. Bourges, *Phys. Rev. B* **78**, 020506 (2008)

[8] M. Campostrini, M. Hasenbusch, A. Pelissetto, P. Rossi and E. Vicari. *Phys. Rev. B* **63**, 214503 (2001)

[9] V. Aji and C. M. Varma. *Phys. Rev. Lett.* **99**, 067003 (2007)

[10] V. Aji and C. M. Varma. *Phys. Rev. B* **79**, 184501 (2009)

[8] V. Balédent, B. Fauqué, Y. Sidis, N. B. Christensen, S. Pailhès, K. Conder, E. Pomjakushina, J. Mesot, and P. Bourges. *Phys. Rev. Lett.* **105**, 027004 (2010)